\documentclass[11pt]{article}
\usepackage{geometry}
\usepackage{mathpazo}

\usepackage{floatrow}
\usepackage{float}
\usepackage{graphics,graphicx}
\usepackage{wrapfig}
\usepackage{amsmath,amssymb}
\usepackage{caption}
\usepackage{subcaption}
\usepackage{sidecap}
\usepackage{tablefootnote}

\usepackage{url}
\usepackage{hyperref}
\usepackage{natbib}
\usepackage{paralist}

\let\olditem\item

\usepackage{etoolbox}
\makeatletter
\patchcmd{\@lbibitem}{\item[\hfil\NAT@anchor{#2}{\NAT@num}]}{\item[\NAT@anchor{#2}{\NAT@num}]}{}{}
\makeatother

\usepackage{xcolor}
\usepackage{tcolorbox}
\usepackage{soul}
\definecolor{darkcyan}{rgb}{0.0, 0.0, 0.7}

\definecolor{boxcolor}{RGB}{193, 229, 255}
\definecolor{figboxcolor}{RGB}{193, 229, 255}
\definecolor{boxcolorgray}{RGB}{227, 227, 227}
\definecolor{darkcyan}{rgb}{0.0, 0.0, 0.7}

\sethlcolor{boxcolorgray}

\usepackage{fancyhdr}
\begin{document}

\noindent
\begin{raggedright}
{\Huge\bfseries The Birth and Early Feedback of Massive Star Clusters Across the Local Universe \\[0.5em]
}\end{raggedright}

\noindent

Jiayi~Sun$^1$,
Rebecca~C.~Levy$^2$,
Cecilia~Bacchini$^3$,
Alberto~D.~Bolatto$^4$,
Alison~H.~Costa$^5$,

Kimberly~Emig$^5$,
Adam~Ginsburg$^6$,
Adam~K.~Leroy$^{7,8}$,
Sean~T.~Linden$^9$,
Elisabeth~A.~C.~Mills$^{10}$,

Dieu~D.~Nguyen$^{11}$,
Miguel~Querejeta$^{12}$,
Eva~Schinnerer$^{13}$,
Thomas~G.~Williams$^{14}$

\vspace{1em}
{
\footnotesize{
\noindent $^1$ Department of Physics and Astronomy, University of Kentucky, 506 Library Drive, Lexington, KY 40506, USA \\
$^2$ Space Telescope Science Institute, 3700 San Martin Drive, Baltimore, MD 21218, USA \\
$^3$ DARK, Niels Bohr Institute, University of Copenhagen, Jagtvej 155, 2200 Copenhagen, Denmark \\
$^4$ Department of Astronomy, University of Maryland, 4296 Stadium Drive, College Park, MD 20742, USA \\
$^5$ National Radio Astronomy Observatory, 520 Edgemont Road, Charlottesville, VA 22903, USA \\
$^6$ Department of Astronomy, University of Florida, P.O. Box 112055, Gainesville, FL 32611, USA \\
$^7$ Department of Astronomy, The Ohio State University, 140 West 18th Avenue, Columbus, Ohio 43210, USA \\
$^8$ Center for Cosmology and Astroparticle Physics, 191 West Woodruff Avenue, Columbus, OH 43210, USA \\
$^9$ School of Physics and Astronomy, Rochester Institute of Technology, 1 Lomb Memorial Dr, Rochester, NY 14623, USA \\
$^{10}$ Department of Physics and Astronomy, University of Kansas, 1251 Wescoe Hall Drive, Lawrence, KS 66045, USA \\
$^{11}$ Department of Astronomy, University of Michigan, 1085 South University Avenue, Ann Arbor, MI 48109, USA \\
$^{12}$ Observatorio Astron\'{o}mico Nacional (IGN), C/Alfonso XII, 3, E-28014 Madrid, Spain \\
$^{13}$ Max-Planck-Institut f\"{u}r Astronomie, K\"{o}nigstuhl 17, D-69117, Heidelberg, Germany \\
$^{14}$ UK ALMA Regional Centre Node, Jodrell Bank Centre for Astrophysics, Department of Physics and Astronomy, The University of Manchester, Oxford Road, Manchester M13 9PL, UK \\
}
}

\begin{tcolorbox}[boxsep=3pt,left=0pt,right=0pt,top=1pt,bottom=1pt,colframe=boxcolor,colback=boxcolor]
\justify
\textbf{Abstract:} 
Young massive star clusters (YMCs; ${\gtrsim}10^4\rm\,M_\odot$) represent a distinct, extreme mode of star formation and the closest present-day analogues of the environments in which globular clusters and Cosmic Noon stellar populations assembled. Given the extreme dust extinction in the earliest phase of YMC formation, (sub-)millimeter observations are the only means of probing their assembly and early feedback. Pioneering works with ALMA are limited to individual region studies and to the most massive YMCs in them, leaving many open questions about the initial conditions of YMC formation, the integrated star and cluster formation efficiencies, the dominant pre-supernova feedback channels, and the prevalence and reach of cluster-driven outflows. ALMA2040 will enable mass-complete, resolved surveys in hundreds of host galaxies, transforming the study of massive cluster formation and feedback from individual case analyses into statistical measurements across the full range of galactic environments.
\end{tcolorbox}

\newpage{}

\vspace{-0.7cm}
\section{Scientific Context and Open Questions}

Massive star clusters (${\gtrsim}10^4\rm\,M_\odot$) form in dense, turbulent environments \citep[$n\,{>}\,10^{3}\rm\,cm^{-3}$;][]{PortegiesZwart_etal_2010,Krumholz_etal_2019}.
Unlike star formation under the more common, lower-density conditions, the massive clusters must form quickly and efficiently convert their natal gas into stars, all while the newly formed stellar population exerts strong feedback on their surroundings \citep{Krumholz_etal_2014}.
Young massive clusters (YMCs) are therefore the hallmark of a distinct, extreme mode of star formation and a natural laboratory for testing how gravity, turbulence, and feedback compete under conditions inaccessible in ordinary local star-forming clouds.
\smallskip

YMCs also provide a critical low-redshift anchor for understanding star and cluster formation across cosmic time. 
At Cosmic Noon ($z\sim1{-}3$), typical gas densities and pressures were comparable to today's most extreme conditions \citep{Tacconi_etal_2020}, and a large fraction of star formation appears to have proceeded in or near massive clusters \citep{Vanzella_etal_2022}.
At higher redshift, essentially all globular clusters observed today formed in the first few Gyrs of cosmic history, yet the conditions and mechanisms that produced them remain poorly constrained \citep{Gratton_etal_2019}.
As resolved observations of forming massive clusters remain out of reach at either Cosmic Noon or higher $z$ \citep[except for rare, strongly lensed objects; e.g.,][]{Adamo_etal_2024}, local analogs with comparable (or at places more extreme) conditions become the best alternative, where one can directly resolve and test the relevant physical processes in YMC formation and feedback.
\smallskip

\begin{figure}[t!]
\centering
\includegraphics[width=0.32\textwidth]{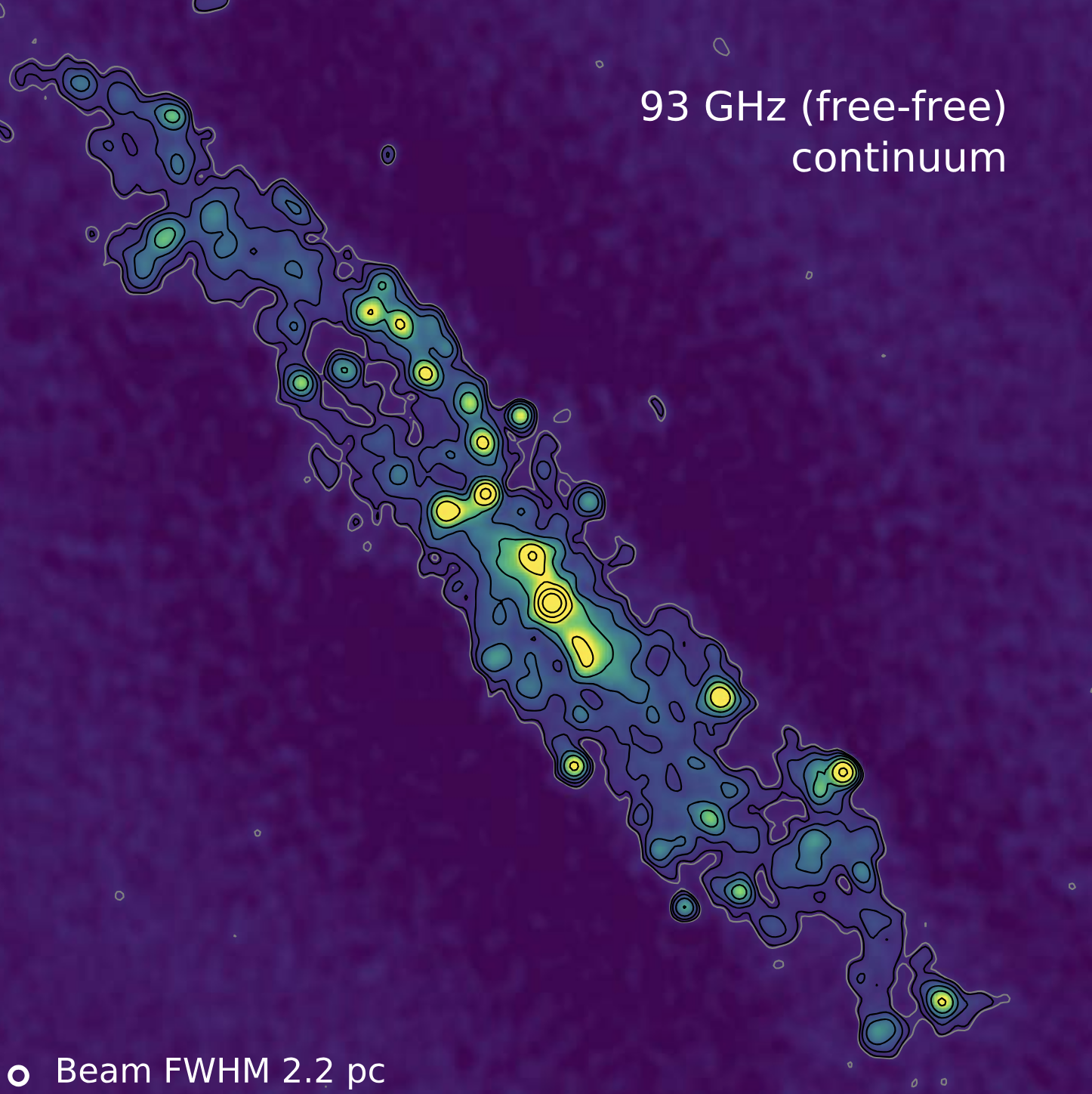}
\includegraphics[width=0.32\textwidth]{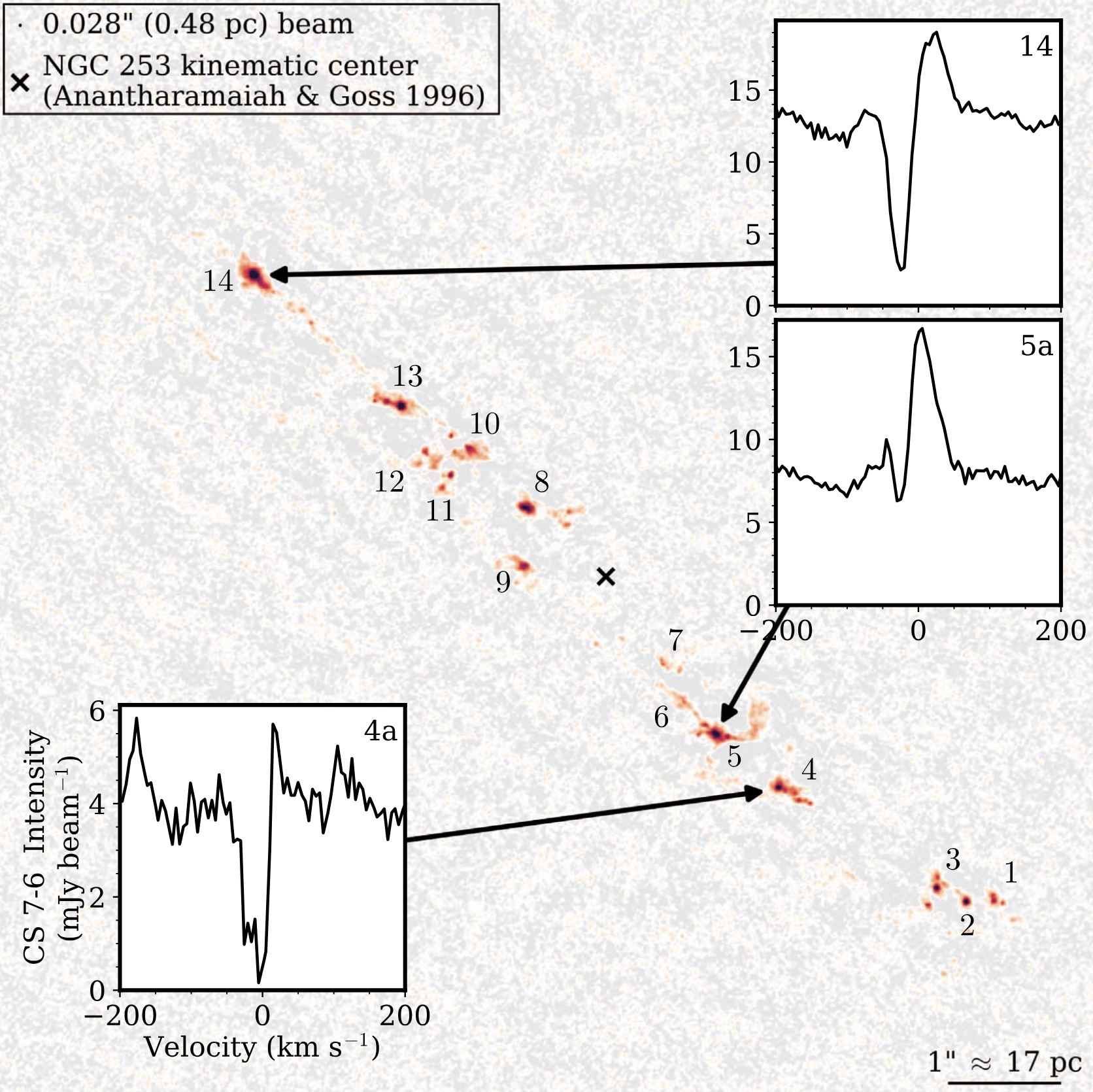}
\includegraphics[width=0.32\textwidth]{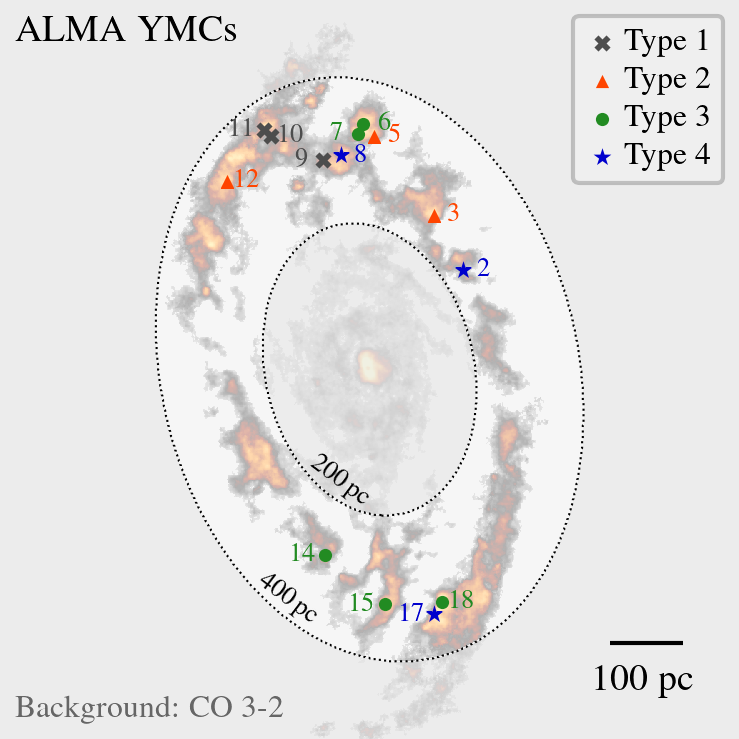}
\caption{
(Sub-)millimeter detections of YMCs with ALMA in the nuclear starburst regions of NGC~4945 \citep[\textit{left panel}; 93~GHz continuum at 2.2~pc resolution;][]{Emig_etal_2020}, NGC~253 \citep[\textit{middle panel}; 350~GHz continuum at 0.5~pc resolution;][]{Levy_etal_2021}, and NGC~3351 \citep[\textit{right panel}; 93 and 350~GHz detections at 8~pc resolution;][]{Sun_etal_2024}.
The middle panel also shows P-Cygni profiles of the high-critical-density, optically thin CS~(7--6) line, indicative of molecular outflows in three YMCs.
The right panel overlays the location of ALMA-identified YMCs on the large-scale gas reservoir traced by CO\,(3--2).
While current instrumental capabilities only allow individual region studies and access to the high-mass tail of YMC populations, ALMA2040 will entirely change the landscape by enabling mass-complete YMC censuses over hundreds of diverse host systems.
}
\label{fig:examples}
\end{figure}

While much of the early work on massive cluster properties was based on optical and near-IR observations \citep[e.g.,][]{Ho_Filippenko_1996,McCrady_etal_2003,Wilson_etal_2006,Kornei_McCrady_2009}, these wavelengths are not effective at probing objects in their earliest stages of formation and feedback due to extreme dust extinction \citep[$A_V \gtrsim 100$~mag;][]{Leroy_etal_2018,Budaiev_etal_2026}.
Instead, (sub-)millimeter observations can trace both the natal gas and dust around these infant massive clusters and their early feedback signatures through thermal dust and free-free emission, molecular lines with different critical densities and optical depths, as well as radio recombination lines.
Recent works with ALMA have begun to exploit this capability by resolving and characterizing YMCs in the Galactic Center and a handful of nearby extragalactic starburst systems \citep[e.g.,][see \autoref{fig:examples}]{Ginsburg_etal_2018,Leroy_etal_2018,Barnes_etal_2019,Emig_etal_2020,Costa_etal_2021,Mills_etal_2021,Rico-Villas_etal_2021,Levy_etal_2021,Levy_etal_2022,Bellocchi_etal_2023,Sun_etal_2024}.
Together, these studies reveal rapid early evolution, efficient gas-to-star conversion, and high cluster formation efficiencies, while also highlighting open questions and unresolved tensions that require a next-generation (sub-)millimeter facility to settle.

\subsection{Open questions in massive star cluster formation}

The sites and initial conditions of massive cluster formation determine much of the eventual outcome, including the final cluster masses and whether they can survive over a long timescale.
Where and how this distinct mode of star formation is triggered appears to vary from system to system.
In nuclear starburst systems, large-scale dynamical driving (e.g., by stellar bars) funnels gas into galaxy centers, where massive clusters can form through either gravitational instabilities or gas collisions \citep[e.g.,][]{Boker_etal_2008,Fukui_etal_2021,Levy_etal_2022}.
In galaxy mergers, tidal compression and shocks can also trigger massive cluster formation, raising both cluster formation efficiency and the upper limit of cluster mass \citep{PortegiesZwart_etal_2010,Adamo_etal_2020}.
Many of these mechanisms, however, have been inferred from or tested against a small number of observed systems, each of which represents only a snapshot at one evolutionary stage \citep[e.g.,][]{Levy_etal_2022,Walker_etal_2024}.
Further progress requires a much larger sample of observed YMCs with a range of host system morphologies and evolutionary stages.
This also demands substantially improved sensitivity and resolution to detect the full YMC population in each system, rather than only the extreme tail accessible with current facilities.
\smallskip

Once the progenitor clouds are formed, their subsequent fragmentation and dynamical evolution strongly shape the size and stellar content of the resulting YMCs.
Proposed scenarios include \textit{in situ} formation, in which individual stars form at approximately the final cluster density; the ``conveyor belt'' model, in which stars are formed in spatially dispersed dense cores which gather as the parent cloud contracts; and the ``popping cluster'' model, in which stars first form in a denser configuration and subsequently expand due to gas expulsion \citep[e.g.,][]{Lada_etal_1984,Longmore_etal_2014}.
These scenarios can be separated by comparing the density and spatial extent of the natal gas with those of the resulting stellar body, as shown by YMC studies in the Galactic Center \citep{Barnes_etal_2019}.
Yet such experiments require resolved, high-dynamic-range dust continuum or molecular line observations probing the structural hierarchy from cloud to core scales, which are limited by small sample size in our Galaxy and remain impractical for the vast majority of extragalactic YMCs (also see a companion white paper by A.~Duarte~Cabral et al.).
\smallskip

Both the star formation efficiency (i.e., the fraction of gas mass converted into stars) and cluster formation efficiency (i.e., the fraction of stellar mass formed in bound clusters) appear to increase in extreme, YMC-forming environments \citep{Krumholz_etal_2019,Adamo_etal_2020}.
Numerical simulations also predict similar trends in the star formation efficiency with gas surface density, though the exact shape and normalization of these relationships remain debated \citep[e.g.,][]{Grudic_etal_2018,Kim_etal_2018,Menon_etal_2023,Polak_etal_2023}.
Most extragalactic measurements to date are limited by incomplete recovery of deeply embedded clusters \citep{Chandar_etal_2017,Rodriguez_etal_2025}, partially due to the limited availability of deep millimeter observations detecting individual YMCs (for $<$10 systems) compared to optical/IR surveys of exposed clusters at similar angular resolution ($>$100 systems).
They also frequently measure the instantaneous ratio of stellar to gas mass, whereas the integrated star formation efficiency must be inferred from an ensemble spanning the full evolutionary sequence.
Conclusive tests of model predictions thus require statistical inference from larger and more complete samples of YMCs, especially in the earliest embedded phase, and ideally distributed over a broad range of galactic environments and gas surface densities.

\subsection{Open questions in massive star cluster feedback}

Stellar feedback disperses the natal gas of forming clusters on relatively short timescales.
Empirical measurements with different methods agree on an overall cluster emergence timescale of a few Myr \citep{Hollyhead_etal_2015,He_etal_2022,Whitmore_etal_2023a,Linden_etal_2024,Sun_etal_2024,Knutas_etal_2025,Wainer_etal_2026}, suggesting that pre-supernova mechanisms are responsible for gas clearing \citep{Chevance_etal_2022}.
The clearing timescale also appears shorter in more massive clusters \citep{Pedrini_etal_2026}, yet the YMC regime remains the least constrained due to limited number statistics and the difficulty of obtaining robust mass and age measurements \citep{Hollyhead_etal_2015,Levy_etal_2024}. 
Establishing the timescale of cluster feedback therefore requires (sub-)millimeter censuses of the starless and earliest assembly phases, combined with existing and future surveys of the exposed cluster population in the same host systems and at similar mass completeness \citep{Sun_etal_2024,Pedrini_etal_2026}.
\smallskip

Aside from the cluster emergence timescale, it is even less clear which specific pre-supernova feedback channels accomplish the gas clearing, and the extreme conditions under which YMCs form can also alter their relative effectiveness \citep{Krumholz_etal_2019}.
Photoionization drives expansion at roughly the ionized gas sound speed, which can fall below the escape velocities of forming YMCs, rendering this channel ineffective for gas removal in dense environments \citep{Bressert_etal_2012,Kim_etal_2018,Levy_etal_2021}.
Stellar winds can reach much higher terminal velocities, but numerical simulations suggest that strong radiative cooling in dense, turbulent media lowers the momentum output of wind-blown bubbles and limits their ability to disperse high surface density gas  \citep{Lancaster_etal_2021,Polak_etal_2023}.
Similarly, radiation pressure is believed to play more important roles at higher surface densities \citep{Skinner_Ostriker_2015,Menon_etal_2023,Menon_etal_2024}, yet many observed YMCs appear to be sub-Eddington to both direct and dust-reprocessed radiation \citep{Leroy_etal_2018,Sun_etal_2024}.
It is also debated whether dust-reprocessed radiation pressure becomes important at very high surface densities and/or very early times  \citep{Olivier_etal_2021,Levy_etal_2021,Menon_etal_2022}.
These considerations together raise the possibility that most feedback channels weaken towards the high-density regime and lead to high star formation efficiencies \citep[e.g.,][]{Krumholz_etal_2019,Dekel_etal_2023}.
Putting this expectation to a definitive test requires robust measurements of the masses, sizes, ages, and feedback signatures of embedded YMCs, which current facilities struggle to measure due to both limited sensitivity and resolution \citep{Levy_etal_2021}.
\smallskip

Gas outflows provide one of the most direct evidence of cluster-scale feedback, with potential imprint on even larger scales \citep{PortegiesZwart_etal_2010,Mills_etal_2021,Nguyen_Thompson_2022}.
Molecular outflows have been detected with optically thin lines in a few very nearby YMCs, with velocities near their escape velocity and possibly high mass loss rates \citep[\autoref{fig:examples} middle panel]{Levy_etal_2021}.
However, the prevalence of such outflows as functions of YMC mass and age has not been established.
Nor is it clear if the cluster-scale feedback also drives gas flows on much larger scales, such as expanding shells or galactic winds \citep{Leaman_etal_2019,Bolatto_etal_2021,Mills_etal_2021,Sheriff_etal_2026}.
Constraining the duty cycle and reach of YMC outflows calls for observations that detect both localized outflow signatures with high-critical-density molecular lines and extended gas and dust distributions in lower-critical-density lines and continuum.
Such measurements remain impractical for most extragalactic YMCs without substantial improvements in both the sensitivity and resolving power of current millimeter facilities.

\section{ALMA2040: Technical Requirements}

We expect ALMA2040 to enable the first systematic census of forming YMCs across hundreds of galaxies in the local universe and revolutionize our understanding of the initial conditions, efficiencies, and impacts of their formation and feedback.
This will be possible with an improvement of 40$\times$ in continuum sensitivity relative to the present-day ALMA, which can be achieved with substantial increases in the number of antennae (200--300), instantaneous bandwidth ($\geq$128~GHz), and additional improvements on the backend facilities.
This massive sensitivity gain will expand the distance limit of YMC studies by sixfold and thus the search volume by a factor of $\sim$250.
It also means routinely reaching high mass completeness of the embedded YMC population instead of probing only its high-mass tail.
Meanwhile, a corresponding 10$\times$ gain in line sensitivity will enable in-depth studies of the multi-scale gas reservoir feeding infant YMCs and experiencing their feedback.
A 3$\times$ larger interferometric baseline (and thus 3$\times$ sharper angular resolution) will lead to resolved individual YMCs and even their internal structures in the nearest host systems, vastly improving current constraints on YMC formation channels and feedback strengths (as they scale strongly with size estimates).
Together, these capabilities would transform the qualitative, individual region studies that currently define the field into robust, statistical surveys spanning the full range of environments in which massive clusters form.

\subsection{Projected capabilities}

ALMA2040 will efficiently survey and characterize infant YMCs in the starless and deeply embedded stages by detecting the thermal emission from their natal dust envelopes at high frequencies (Bands 7--10). At 50~Mpc, YMCs with a gas mass of $10^4\rm\,M_\odot$ would have continuum flux densities of $\sim$8~$\mu$Jy at 350~GHz (modulo assumptions on dust temperature, opacity, and dust-to-gas ratio; \citealp{Leroy_etal_2018}), which ALMA2040 can detect at $\rm S/N=10$ with merely 1~hour of integration.
ALMA2040 can also marginally resolve such YMCs (with typical sizes $\sim$1~pc) out to the same distance, thanks to its improved maximum baseline, corresponding to $\sim$4~mas resolution at 350~GHz.
Therefore, ALMA2040 will be capable of conducting mass-complete and resolved surveys of YMCs across hundreds of host systems within $50$~Mpc, which brings a large number of merger-driven starbursts and other rare, YMC-forming systems within its reach.
\smallskip

YMCs with already formed massive stars can also be selected by free-free emission from the locally ionized gas reservoir.
At the same 50~Mpc distance, YMCs with a young, $\lesssim$1~Myr old stellar population of $10^4\rm\,M_\odot$ would produce free-free continuum at $\sim$1~$\mu$Jy (modulo assumptions on gas temperature; \citealp{Emig_etal_2020}), which is recoverable at $\rm S/N>3$ with 1~hour integration on ALMA2040 in Bands 1--3.
A $\geq$128~GHz instantaneous bandwidth with ALMA2040 will also enable strong constraints on the spectral slope and decontamination of the low-frequency continuum by potential synchrotron emission in older clusters.
Therefore, the combination of deep continuum observations at high and low frequencies, possibly done with simultaneous multi-band technologies, can yield highly complete YMC surveys with empirical markers of evolutionary stages.
\smallskip

Future YMC studies can also benefit from a 10$\times$ increase in line sensitivity with ALMA2040.
Observations of $^{12}$CO, CO isotopologues, and higher-critical density molecular lines with multiple array configurations can map the hierarchical gas structures around and within the nearest extragalactic YMCs (see white paper by A.~Duarte~Cabral et al.), complementing and contrasting with similar observations in the Galactic Center \citep[e.g.,][]{Walker_etal_2026}.
Fainter, optically thin tracers of the densest gas can also reveal signatures of localized outflows \citep{Levy_etal_2021}, and ALMA2040 will facilitate such detections in many more galaxies out to at least 3$\times$ farther distances.
Last but not least, an instantaneous bandwidth of $\geq$128~GHz (16$\times$ improvement over the current ALMA and 4--8$\times$ over the Wideband Sensitivity Upgrade) also means vastly improved ``spectral grasp'' and efficient multiplexing for all spectral line detections.
It will especially enable spectral stacking of fainter lines, including radio recombination lines, which further enrich the inventory of multiphase gas tracers and measurements of YMC feedback with ionized gas content, temperature, and possibly even kinematics \citep[e.g.,][]{Mills_etal_2021,Emig_etal_2026}.

\begin{figure}[h!]
\centering
\includegraphics[width=\textwidth]{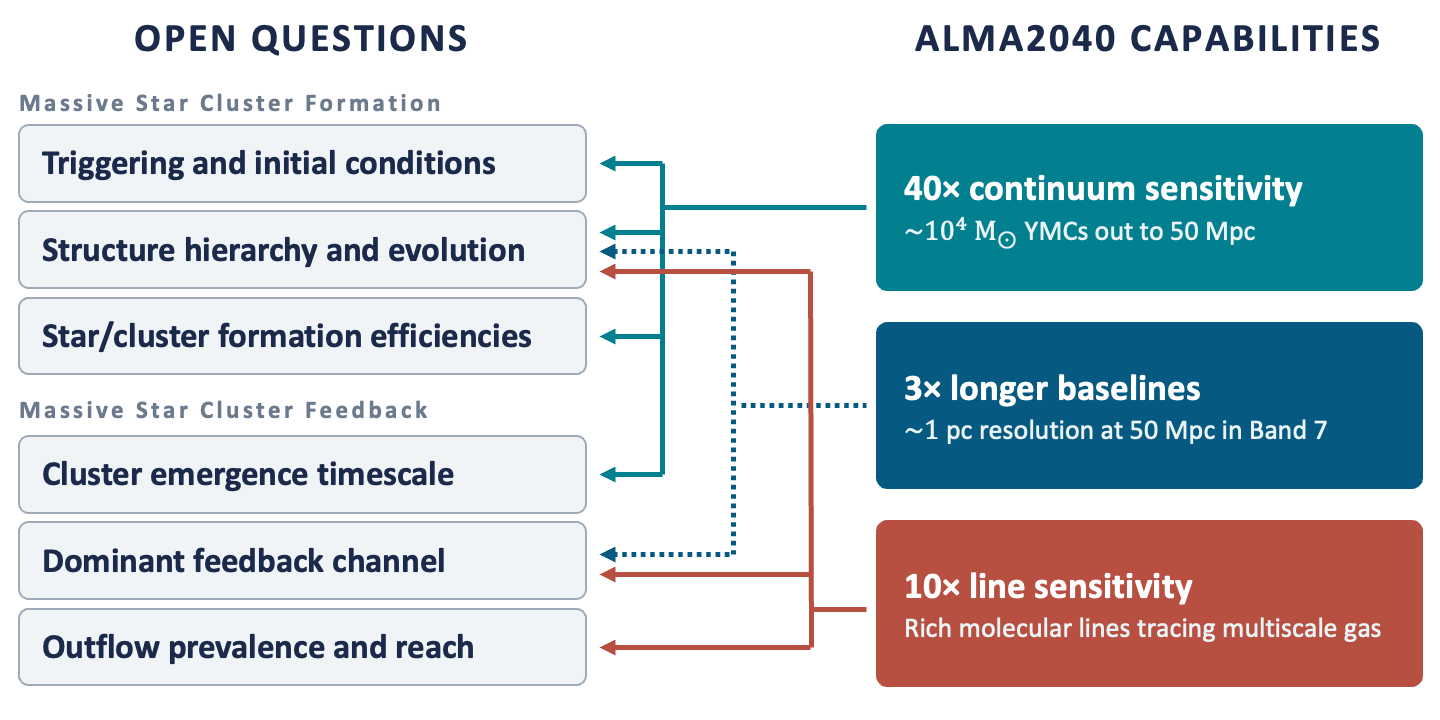}
\vspace{-1.5\baselineskip}
\caption{
A summary of the open questions on YMC formation and feedback highlighted in this white paper and which ALMA2040 capabilities are critical for addressing each of them.
}
\label{fig:schematic}
\end{figure}

\section{Synergies with Current and Future Facilities}

This science will also be boosted by complementary programs with many existing and planned future observatories.
For example, \textit{JWST} and \textit{Roman} will provide direct access to the stellar content of exposed and moderately embedded YMCs across a similarly large sample of host galaxies, complementing the (sub-)millimeter census of earlier evolutionary stages and enabling robust measurements of cluster mass, age, formation efficiency, and survival rate.
Dedicated follow-up programs on ELT and GMT will further access stellar kinematics, cluster dynamical masses, and chemical abundances, which are essential for verifying connections between present-day YMCs and globular clusters.
SKA and the ngVLA will not only extend continuum and line measurements to lower frequencies, but also resolve the substructures within individual clusters thanks to their longer interferometric baselines.
These will allow us to fully capitalize on the revolutionary (sub-)millimeter capabilities brought by ALMA2040.

\section*{Acknowledgments}
This work was developed as part of the broader ALMA2040 community effort. We acknowledge the Lorentz Center and NOVA for supporting the community workshops and coordination activities that helped initiate and shape the ALMA2040 White Paper process.

\bibliography{main}

\end{document}